# Special Theory of Super Relativity and a Possibility to Exceed the Speed of Light

V. I. Klyukhin

*Abstract*—Modification of special theory of relativity is proposed to describe the propagation of signals with superluminal velocity. Modified kinematics and Lorentz transformations of Maxwell's equations are described. A possible experiment on measuring the speed of light for verification of the theory is discussed. It is shown that replacing the maximum speed in special theory of relativity by the neutrino superluminal velocity measured by the OPERA collaboration leads to lower value of the muon mass reconstructed from the muon decay products with the modified relativistic kinematics. This value is in strong conflict with the precise muon mass measurements that excludes introducing the superluminal velocity of signals into special theory of relativity.

## I. Introduction

TWO great masterworks influenced on this study: a famous article "On the Electrodynamics of moving bodies" [1] by Albert Einstein and a perfectly written book "Ryman geometry and tensor analysis" [2] by Petr K. Rashevsky.

In this study we follow the conjecture, expressed by A. Einstein, "that not only in mechanics, but in electrodynamics as well, the phenomena do not have any properties corresponding to the concept of absolute rest, but that in all coordinate systems in which the mechanical equations are valid, also the same electrodynamic and optical laws are valid, as has already been shown *for quantities of the first order*" [1] (the first Einstein's postulate (A) or "principle of relativity"). In the same time we do not follow the second Einstein's postulate of independence of the light speed in empty space of whether the ray of light is emitted by a body at rest or a body in motion. We modify this postulate in the way as follows: (B) there is an interaction between some objects in empty space that produces a "super" light that is always propagated with a definite velocity $S$ which exceeds the speed of light and is independent of state of motion of the emitted body (*a principle of existence of the constant superluminal velocity of the interaction propagation*).

Combining these principles A and B we modify the special theory of relativity to allow the propagation of *super light* with a velocity exceeded the light speed in empty space.

In Section II we obtain the modified Lorentz transformations of the space-time coordinates of the events when the transition from one inertial frame of reference to another one occurs. In Section III we analyze the modified Lorentz transformations and apply it to kinematics. In Section IV we transform the components of the electromagnetic field to a moving body. In Section V we discuss an experiment with two artificial earth satellites to measure the light speed in empty space and to verify the special theory of super relativity. Finally, in Section VI we describe the application of the theory to the OPERA collaboration data [3]. We draw the conclusion from the study in Section VII.

We use *SI units* throughout all the study except of Section IV where we use *Gaussian System of equations* and Section VI where we use *natural units* with speed of light equal to unity.

## II. Modified Lorentz Transformations

Consider two Cartesian coordinate systems in a lab system in which the Newtonian mechanical equations are valid. Let the $X$-axes of the two systems coincide and their $Y$- and $Z$-axes be parallel. Let each system be supplied with a rigid measuring rod and a number of clocks and the two measuring rods and all the clocks of the two systems be exactly alike that means *the units of length or the units of time in both systems we assume exactly the same*. Let the origins of two systems coincide at the moment of time equal to zero in both coordinate systems.

The origin of one of the two systems ($K'$) shall now be imparted a constant velocity $v$ in the direction of increasing $x$ of the other system ($K$), which is at rest in the lab system, and we assume that in each time $t$ the axes of the moving system are parallel to the axes of the system at rest.

Let the space be measured both from the system at rest $K$ by means of the measuring rod at rest and from the moving system $K'$ by means of the measuring rod moving along with it, and the coordinates $x$, $y$, $z$ and $\xi$, $\eta$, $\zeta$ are obtained in this way.

We now imagine that there are detectors mounted at equal distances $d$ from the origin on the axes of the moving system $K'$ at the coordinate values $\xi = \eta = \zeta = d$ and these detectors can register a signal from some interaction occurred in the space point where the origins of both systems $K$ and $K'$ coincided at the momentum of time $t = 0$. Let call this signal a

---

ray of *super light* and let it propagate with a superluminal velocity *S* which is independent of state of motion of the emitted body, i.e. velocity of such a signal measured in the lab system *K* and in the moving system *K'* *has the same numerical value*.

We denote this velocity *S* like a distance *d* passed by the ray of super light divided by a time interval *τ* needed for that in the system at rest. In this way

$$S = \frac{\xi}{\tau} = \frac{\eta}{\tau} = \frac{\zeta}{\tau} = \frac{d}{\tau}, \qquad (1)$$

and *τ* also is a time interval that the ray of super light needs to be registered by the detectors in the system, where they are at rest, i. e. in the system *K'*.

In Fig. 1 we draw the propagation of super light in the systems *K* and *K'* at different moments of time *t*, starting from *t* = 0, in a case when *v* = 0.01·*S*. For simplicity we display only two coordinate axes of both coordinate systems.

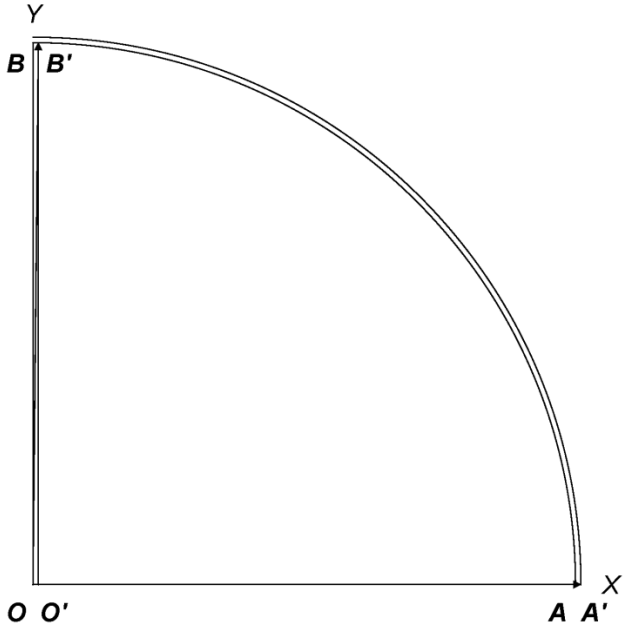

Fig. 1. Propagation of super light with respect to the coordinate system at rest *K* and the moving system *K'* in the case of *v* = 0.01·*S*. The position of the moving system *K'* is shown at the time interval *t* = *τ*.

Consider the propagation of super light in the lab system *K* from an origin *O* to the detectors *A* and *B* which move together with the system *K'* with velocity *v* and sequentially change their positions from a point *A* to points *A'*, *A''*, etc, and from a point *B* to points *B'*, *B''*, etc. The origin of the system *K'* moves from the point *O* to points *O'*, *O''*, etc in the same moments of time *t*.

Determine in system *K* a time interval $t_x$ that super light originated from *O* needs to be registered by the detector *A*. When the ray of super light passes the distance *OA* = *d* for a time interval $t_1 = \tau = d/S$, the detector *A* passes the distance *AA'* = *τv*, and to reach this position, the ray of super light needs additional time $t_2 = \tau v/S$. When the ray of super light passes the distance *AA'* for a time interval $t_2$, the detector *A* passes the distance $A'A'' = \tau v^2/S$, and to reach new position of the detector, the ray of super light needs an additional time interval $t_3 = \tau v^2/S^2$, etc. Thus, we can express time $t_x$ as a sequence of time intervals

$$t_x = t_1 + t_2 + t_3 + t_4 + \ldots = \tau\left(1 + \frac{v}{S} + \frac{v^2}{S^2} + \frac{v^3}{S^3} + \ldots\right) =$$

$$\tau\left[1 + \frac{v}{S} + \left(1 + \frac{v}{S}\right)\frac{v^2}{S^2} + \ldots\right] = \tau\left(1 + \frac{v}{S}\right)\left(1 + \frac{v^2}{S^2} + \ldots\right) = \quad (2)$$

$$\tau\left(1 + \frac{v}{S}\right) \Big/ \left(1 - \frac{v^2}{S^2}\right).$$

When we use (1) and assume $x = S \cdot t_x$, then

$$t_x = \left(\tau + \frac{v}{S^2}\xi\right) \Big/ \left(1 - \frac{v^2}{S^2}\right), \qquad (3)$$

$$x = (\xi + v\tau) \Big/ \left(1 - \frac{v^2}{S^2}\right). \qquad (4)$$

Determine time $t_y$ that super light originated from *O* needs to be registered by the detector *B*. When the ray of super light passes into direction of *B* with velocity *S*, it goes along the *Y*-axis with velocity $\sqrt{S^2 - v^2}$. Thus, the ray of super light reaches the detector *B* in a time interval

$$t_y = \eta \Big/ \sqrt{S^2 - v^2} = \tau \Big/ \sqrt{1 - \frac{v^2}{S^2}}, \qquad (5)$$

when we apply (1). A similar expression we obtain for a time interval $t_z$ that super light needs to reach the detector *C* located at the distance *d* from the origin of the system *K'* on the *Z*-axis:

$$t_z = \zeta \Big/ \sqrt{S^2 - v^2} = \tau \Big/ \sqrt{1 - \frac{v^2}{S^2}} = t_y. \qquad (6)$$

Both time intervals $t_y$ and $t_z$ differ from $t_x$. From (1), (2), (5), and (6) we obtain the ratios

$$t_x = \tau \Big/ \left(1 - \frac{v}{S}\right) = \xi/(S - v), \qquad (7)$$

$$t_x/t_y = t_x/t_z = \sqrt{1 + \frac{v}{S}} \Big/ \sqrt{1 - \frac{v}{S}}. \qquad (8)$$

From (5) − (8) we derive that $t_x = t_y = t_z$ only in two cases: first, when $\xi = \eta = \zeta = 0$, and second, when *v* = 0. Thus, to keep the equality of those three time intervals we can apply two good approximations: first, *τ* → 0, and second, *S* → ∞.

A. Einstein used in [1] the first approximation, considering an infinitesimal propagation of a signal assuming its velocity to be equal to the velocity of propagation of electromagnetic interaction in empty space, *c* = 3·10⁸ m/s. In that case, if we assume the velocity *v* to be equal to the orbital velocity of the

Earth $3 \cdot 10^4$ m/s, we obtain $t_x/t_y = t_x/t_z \approx 1.0001$, $t_x/\tau \approx 1.0001$, and $t_y/\tau = t_z/\tau \approx 1.000000005$.

In this study we use the second approximation $S \to \infty$. First, we change *the units* of time and length in the moving system $K'$ in a way as follows:

$$t' = \gamma\tau, \quad x' = \gamma\xi, \quad y' = \gamma\eta, \quad z' = \gamma\zeta, \qquad (9)$$

where

$$\gamma = 1 \Big/ \sqrt{1 - \frac{v^2}{S^2}} \qquad (10)$$

is modified Lorentz factor.

These transformations keep the numerical value of the velocity of super light in the moving system $K'$ unchanged:

$$S = \frac{x'}{t'} = \frac{y'}{t'} = \frac{z'}{t'}. \qquad (11)$$

Transformations (9) and equations (3), (5), and (6) give the ratios

$$tx = \gamma\left(t' + \frac{v}{S^2}x'\right), \quad t_y = t', \quad t_z = t'. \qquad (12)$$

If we apply the principle of the constancy of the velocity of super light $S$ in any system at rest, in particular in system $K$, we obtain the ratios between the coordinates of two systems $K$ and $K'$ as follows:

$$x = St_x = \gamma(x' + vt'), \quad y = St_y = y', \quad z = St_z = z'. \qquad (13)$$

To obtain the transformations of the coordinates and time when the transition from the moving system $K'$ to the system at rest $K$ occurs, we combine $x$, $y$, $z$, and $t_x$ from (12) and (13) in a quadratic form

$$s^2 = -S^2 t_x^2 + x^2 + y^2 + z^2. \qquad (14)$$

This form is invariant with respect to both systems $K$ and $K'$:

$$s^2 = -S^2 t_x^2 + x^2 + y^2 + z^2 =$$
$$\left(-S^2 t_x'^2 + x'^2\right)\left(1 - \frac{v^2}{S^2}\right) \Big/ \left(1 - \frac{v^2}{S^2}\right) + y'^2 + z'^2 = \qquad (15)$$
$$-S^2 t'^2 + x'^2 + y'^2 + z'^2.$$

If $s = 0$, then the coordinates $x$, $y$, and $z$ in the lab system $K$ lie on the sphere of radius $S \cdot t_x$ whose center is located at the origin of the coordinate system $K$ and the coordinates $x'$, $y'$, and $z'$ in the moving system $K'$ lie on the sphere of radius $S \cdot t'$ whose center is located at the origin of the coordinate system $K'$. The time interval $t_x$ plays the same role in the lab system $K$ that the time interval $t'$ plays in the moving system $K'$. Let call $t_x$ a proper time in the system at rest $K$ and denote it $t$.

Thus, when we obey the principles (A) and (B) formulated in Section I, we obtain the modified Lorentz transformations of the space-time coordinates of the events, when the transition from one inertial frame of reference to another one occurs:

$$t = \gamma\left(t' + \frac{v}{S^2}x'\right), \quad x = \gamma(x' + vt'), \quad y = y', \quad z = z'. \qquad (16)$$

$$t' = \gamma\left(t - \frac{v}{S^2}x\right), \quad x' = \gamma(x - vt), \quad y' = y, \quad z' = z. \qquad (17)$$

Here the primed coordinate $x'$, $y'$, and $z'$ and time $t'$ are measured in the system $K'$ moving with the velocity $v$ with respect to the lab system $K$, and the unprimed coordinates $x$, $y$, and $z$ and time $t$ are measured in the system at rest $K$.

In the modified Lorentz transformations (16) and (17) the speed of light does not play an exceptional role anymore and becomes a velocity of propagation of one of the possible interactions. An exclusive velocity is the velocity of the propagation of super light $S$ that has the same numerical value in all inertial frames of reference, in particular, in the systems $K$ and $K'$, and could exceed the velocity of the electromagnetic propagation. The propagation of the weak, strong, gravitational or any unknown interaction could be considered as the super light propagation.

### III. SUPER RELATIVISTIC KINEMATICS AND DYNAMICS

#### A. Analysis of the Modified Lorentz Transformations

To obtain the modified Lorentz transformations (16) we scaled *the units* of time and length in the moving system $K'$ according to (9):

$$[t'] = \gamma[\tau], \quad [x'] = \gamma[\xi], \quad [y'] = \gamma[\eta], \quad [z'] = \gamma[\zeta],$$
$$[\tau] = [t], \quad [\xi] = [x], \quad [\eta] = [y], \quad [\zeta] = [z], \qquad (18)$$

where the brackets are used to denote *the units* of variables.

A. Einstein used this scaling of *the units* implicitly [1], considering the differential equations and adding an additional factor $a$ that compensated Lorentz factor $\gamma$ in the equations of type (3) − (6).

Consider a rigid rod that lies along the $X'$-axis of the coordinate system $K'$ and moves together with it with velocity $v$ with respect to the lab system $K$. Let us measure the length of this rod in the system at rest $K$. To perform this measurement we have to determine the coordinates $x_1$ and $x_2$ in the same moment of time $t$. According to (17), we express the abscissa of the rod edges in the system $K'$ as follows:

$$x_1' = \gamma(x_1 - vt), \quad x_2' = \gamma(x_2 - vt),$$
$$x_2' - x_1' = \gamma(x_2 - x_1). \qquad (19)$$

It seems the length of the rod in the system $K$ is contracted with respect to the length of this rod in the system $K'$. If we remind that to obtain the transformations (17) we scaled *the unit* of length in $K'$ in comparison with this *unit* in $K$ according to (9) and (18), we get

$$x_2' - x_1' = \gamma(\xi_2 - \xi_1), \quad x_2 - x_1 = \xi_2 - \xi_1, \qquad (20)$$

thus, in the same *units* in both systems the length of the rod does not change.

Consider two different events occurred in the same point $x'$ of the moving system $K'$ in two different moments $t_1'$ and $t_2'$. According to (16), we express the corresponded time intervals $t_1$, $t_2$ and $t_2 - t_1$ in the system $K$ as follows:

$$t_1 = \gamma\left(t_1' - \frac{v}{S^2}x'\right), \quad t_2 = \gamma\left(t_2' - \frac{v}{S^2}x'\right),$$
$$t_2 - t_1 = \gamma(t_2' - t_1'). \tag{21}$$

It seems the time interval in the system $K$ is longer than the time interval between two events occurred in the same point of the system $K'$. If we remind that to obtain the transformations (16) we scaled *the unit* of time in $K'$ in comparison with this unit in $K$ according to (9) and (18), we obtain

$$\gamma(t_2' - t_1') = \tau_2 - \tau_1, \quad t_2 - t_1 = \tau_2 - \tau_1, \tag{22}$$

thus, in the same *units* in both systems the time interval does not change.

### B. Addition of Velocities

In the system $K'$ moving with velocity $v$ along the $X$-axis of the system $K$ let there be a point moving with a velocity $u'$ according to the equations

$$u_x' = \frac{dx'}{dt'}, \quad u_y' = \frac{dy'}{dt'}, \quad u_z' = \frac{dz'}{dt'}. \tag{23}$$

Differentiating the formulas (16) we obtain

$$dt = \gamma\left(dt' + \frac{v}{S^2}dx'\right),$$
$$dx = \gamma(vdt' + dx'), \quad dy = dy', \quad dz = dz'. \tag{24}$$

From these expressions we can obtain the components of the point velocity on the axes $X$, $Y$, and $Z$ of the system $K$ as follows:

$$u_x = \frac{dx}{dt} = \left(v + \frac{dx'}{dt'}\right)\bigg/\left(1 + \frac{v}{S^2}\frac{dx'}{dt'}\right),$$
$$u_y = \frac{dy}{dt} = \frac{dy'}{\gamma dt'}\bigg/\left(1 + \frac{v}{S^2}\frac{dx'}{dt'}\right), \tag{25}$$
$$u_z = \frac{dz}{dt} = \frac{dz'}{\gamma dt'}\bigg/\left(1 + \frac{v}{S^2}\frac{dx'}{dt'}\right).$$

Finally we get

$$u_x = (v + u_x')\bigg/\left(1 + \frac{v}{S^2}u_x'\right),$$
$$u_y = \frac{u_y'}{\gamma}\bigg/\left(1 + \frac{v}{S^2}u_x'\right), \quad u_z = \frac{u_z'}{\gamma}\bigg/\left(1 + \frac{v}{S^2}u_z'\right). \tag{26}$$

Consider a special case when $u_x'$ is the velocity of propagation of the electromagnetic interaction in empty space, $c = 299792458$ m/s, and $v = 1.6\cdot 10^4$ m/s. Let the ray of light is emitted in the direction of the movement of the system $K'$, then

$$c_x^+ = (v + c)\bigg/\left(1 + \frac{vc}{S^2}\right), \quad c_y = c_z = 0. \tag{27}$$

If the ray of light is emitted in the direction opposite to the movement of the system $K'$, then

$$c_x^- = (v - c)\bigg/\left(1 - \frac{vc}{S^2}\right), \quad c_y = c_z = 0. \tag{28}$$

The numerical value of $c_x^+$ is related to the numerical value of $c_x^-$ by

$$\frac{|c_x^+|}{|c_x^-|} = \frac{c+v}{c-v} = 1.0001, \tag{29}$$

when $S \to \infty$. The ratio of the numerical values $c_x^+$ to $c_x^-$ is equal to 1.00008 when $S = 2\cdot c$.

### C. Geometrical Interpretation of the Modified Lorentz Transformations

From geometrical point of view equations (15) describe a 4-vector **x** in 4-dimensional pseudo Euclidian space of index 1 [2]. In orthonormal coordinate system $x^0$, $x^1$, $x^2$, $x^3$ the scalar square of this 4-vector is

$$\mathbf{x}^2 = -x^{0^2} + x^{1^2} + x^{2^2} + x^{3^2}, \tag{30}$$

where

$$x^0 = St, \quad x^1 = x, \quad x^2 = y, \quad x^3 = z. \tag{31}$$

In this sense the modified Lorentz transformations (16) and (17) describe the rotation of the 4-vector **x** in 4-dimensional pseudo Euclidian space of index 1 that gives the transformation of coordinates $x^0$, $x^1$, $x^2$, $x^3$ to the coordinates

$$x^{0'} = St', \quad x^{1'} = x', \quad x^{2'} = y', \quad x^{3'} = z'. \tag{32}$$

The transition from one inertial frame of reference, $K$, to another one, $K'$, is equivalent to the transition from one orthonormal coordinate system in 4-space to another one. This transition could be described in a general case of the 4-vector **x** rotation and translation with a formula

$$x^{i'} = A_i^{i'} x^i + A^{i'}, \quad (i = 0, 1, 2, 3), \tag{33}$$

where we assume a summation over twice appeared suffices. The pseudo orthonormal matrix $A_i^{i'}$ of order 4 index 1 is related with its inverse matrix $A_{i'}^{i}$ by [2]

$$\begin{Vmatrix} A_{0'}^{0} & A_{0'}^{1} & A_{0'}^{2} & A_{0'}^{3} \\ A_{1'}^{0} & A_{1'}^{1} & A_{1'}^{2} & A_{1'}^{3} \\ A_{2'}^{0} & A_{2'}^{1} & A_{2'}^{2} & A_{2'}^{3} \\ A_{3'}^{0} & A_{3'}^{1} & A_{3'}^{2} & A_{3'}^{3} \end{Vmatrix} = \begin{Vmatrix} A_{0}^{0'} & -A_{1}^{0'} & -A_{2}^{0'} & -A_{3}^{0'} \\ -A_{0}^{1'} & A_{1}^{1'} & A_{2}^{1'} & A_{3}^{1'} \\ -A_{0}^{2'} & A_{1}^{2'} & A_{2}^{2'} & A_{3}^{2'} \\ -A_{0}^{3'} & A_{1}^{3'} & A_{2}^{3'} & A_{3}^{3'} \end{Vmatrix} \tag{34}$$

In the case of the modified Lorentz transformations (17), that we rewrite in the form

$$x^{0'} = \gamma\left(x^0 - x^1 \frac{v}{S}\right), \quad x^{1'} = \gamma\left(-x^0 \frac{v}{S} + x^1\right),$$
$$x^{2'} = x^2, \quad x^{3'} = x^3,$$
(35)

the inverse matrix $A^i_{i'}$ has a form

$$\begin{Vmatrix} \gamma & \gamma\frac{v}{S} & 0 & 0 \\ \gamma\frac{v}{S} & \gamma & 0 & 0 \\ 0 & 0 & 1 & 0 \\ 0 & 0 & 0 & 1 \end{Vmatrix}$$
(36)

A curve in pseudo Euclidian 4-space, when depends from one parameter, e. g. $t = x^0/S$, describes the movement of particle in any inertial frame of reference [2]. If the velocity of the particle is less than velocity of super light $S$, then 4-trajectory of this particle lies inside an isotropic hyper cone and has an imaginary length. In each point of this 4-trajectory we can construct an imaginary unit tangential 4-vector

$$\boldsymbol{\tau} = \frac{d\mathbf{x}}{d\sigma}, \quad \boldsymbol{\tau}^2 = -1, \quad \sigma = \frac{s}{i},$$
(37)

where $s = s(t)$ is imaginary length of the curve, $i$ is imaginary unit, and $\sigma$ is a real parameter.

The coordinates of 4-vector $\boldsymbol{\tau}$ can be expressed in the form [2]

$$\tau^i = \frac{dx^i}{\sqrt{dx^{0^2} - dx^{1^2} - dx^{2^2} - dx^{3^2}}}, \quad (i = 0, 1, 2, 3).$$
(38)

### D. Super Relativistic Dynamics of Particle

Consider the particle moving in the inertial system at rest $K$ with a velocity $v$. In this case we can tie to the particle a local inertial system $K'$ moving with respect to the system at rest $K$ with the same velocity $v$ and apply all the formalism, developed above, to the transformation of variables, attached with the particle, when the transition from the one system to another occurs.

If we assume the dependence of mass $m$ of the particle from the velocity $v$ in a form

$$m = \gamma m_0,$$
(39)

where $m_0$ is a mass of the particle at rest and $S$ is a velocity of super light, then the total energy $E$ of the particle is related to its mass $m$ by

$$E = mS^2,$$
(40)

and the energy of the particle at rest $E_0$ is given by

$$E_0 = m_0 S^2.$$
(41)

In each point of 4-trajectory of the particle in pseudo Euclidian 4-space we construct a 4-momentum $E_0\boldsymbol{\tau}$ of the particle in a way as follows:

$$E_0\tau^0 = mS^2, \quad E_0\tau^1 = mv_x S,$$
$$E_0\tau^2 = mv_y S, \quad E_0\tau^3 = mv_z S.$$
(42)

where

$$v_x = \frac{dx}{dt}, \quad v_y = \frac{dy}{dt}, \quad v_z = \frac{dz}{dt}.$$
(43)

The second Newton's law we can write then in the form

$$\mathbf{F} = \frac{d(m\mathbf{v})}{dt},$$
(44)

i. e. a force $\mathbf{F}$ is equal to the derivative of momentum $m\mathbf{v}$ with respect to time $t$, where mass $m$ is not a constant that gives an additional term in the differentiation.

### IV. TRANSFORMATION OF THE ELECTROMAGNETIC FIELD COMPONENTS

Follow to P. K. Rashevsky [2] consider a particle of charge $e$ moving with a velocity $\mathbf{v}$ in the electromagnetic field at the system at rest $K$. Let $\mathbf{E}$ ($E_x$, $E_y$, $E_z$) and $\mathbf{H}$ ($H_x$, $H_y$, $H_z$) be electric and magnetic field strength, respectively. In this field the particle experiences an action of Lorentz force

$$\mathbf{F} = e\mathbf{E} + \frac{e}{c}[\mathbf{v}\mathbf{H}],$$
(45)

where $c$ is the speed of propagation of electromagnetic interaction in empty space measured at system at rest $K$.

According to the *principle of relativity* (A) we assume this force is valid in any inertial frame of reference. From (44) and (45) we derive an equation

$$\frac{d(m\mathbf{v})}{dt} = e\left\{\mathbf{E} + \frac{1}{c}[\mathbf{v}\mathbf{H}]\right\},$$
(46)

that we write in the form of the vectors components:

$$\frac{d(mv_x)}{dt} = e\left\{E_x + \frac{1}{c}(v_y H_z - v_z H_y)\right\},$$
$$\frac{d(mv_y)}{dt} = e\left\{E_y + \frac{1}{c}(v_z H_x - v_x H_z)\right\},$$
$$\frac{d(mv_z)}{dt} = e\left\{E_z + \frac{1}{c}(v_x H_y - v_y H_x)\right\}.$$
(47)

If we multiply the terms of equations (47) by $S \cdot dt$ and take into account equations (42), (43), and (31) then we obtain

$$d(E_0\tau^1) = e\left\{E_x dx^0 + \frac{S}{c}(H_z dx^2 - H_y dx^3)\right\},$$
$$d(E_0\tau^2) = e\left\{E_y dx^0 + \frac{S}{c}(-H_z dx^1 + H_x dx^3)\right\},$$
$$d(E_0\tau^3) = e\left\{E_z dx^0 + \frac{S}{c}(H_y dx^1 - H_x dx^2)\right\}.$$
(48)

To these three formulas we add the forth equation that expresses the differential of the particle energy $mS^2$ which is

equal to the elementary work performed with the particle by the electric field strength:

$$d(E_0\tau^0) = e\{E_x dx^1 + E_x dx^2 + E_x dx^3\}. \quad (49)$$

Finally we turn the coordinates of 4-vector $E_0\tau$ into covariant form and write equations (49) and (48) as follows:

$$d(E_0\tau_i) = eF_{ij}dx^j, \quad (50)$$

where antisymmetric tensor $F_{ij} = -F_{ji}$ has a form

$$\begin{Vmatrix} F_{00} & F_{01} & F_{02} & F_{03} \\ F_{10} & F_{11} & F_{12} & F_{13} \\ F_{20} & F_{21} & F_{22} & F_{23} \\ F_{30} & F_{31} & F_{32} & F_{33} \end{Vmatrix} = \begin{Vmatrix} 0 & -E_x & -E_y & -E_z \\ E_x & 0 & \frac{S}{c}H_z & -\frac{S}{c}H_y \\ E_y & -\frac{S}{c}H_z & 0 & \frac{S}{c}H_x \\ E_z & \frac{S}{c}H_y & -\frac{S}{c}H_x & 0 \end{Vmatrix}. \quad (51)$$

This tensor permits to transform the components of the electric and magnetic fields from the lab system $K$ to the moving system $K'$ with a tensor transformation low as follows:

$$F_{i'j'} = A_{i'}^i A_{j'}^j F_{ij}, \quad (i, j = 0, 1, 2, 3), \quad (52)$$

where matrix $A_{i'}^i$ describes the transformation of the coordinates (33) and has the form (36).

Thus, the components of the electric and magnetic fields viewed from a frame moving with velocity $v$ with respect to the lab system are given by

$$E'_x = E_x, \quad E'_y = \gamma\left(E_y - \frac{v}{c}H_z\right),$$
$$E'_z = \gamma\left(E_z + \frac{v}{c}H_y\right), \quad (53)$$

$$H'_x = H_x, \quad H'_y = \gamma\left(H_y + \frac{vc}{S^2}E_z\right),$$
$$H'_z = \gamma\left(H_z - \frac{vc}{S^2}E_y\right). \quad (54)$$

To make Maxwell's equations invariant in all the inertial frames of reference the velocity of super light $S$ in transformations (53) and (54) should not differ much from the speed of light $c$ in empty space. To estimate the value of the velocity of super light we consider transformations (53) and (54) in a case when $v \ll S$. In this case we can write the transformations (53) and (54) in the vector form

$$\mathbf{E}' = \mathbf{E} + \frac{1}{c}[\mathbf{vH}], \quad \mathbf{H}' = \mathbf{H} - \frac{c}{S^2}[\mathbf{vE}]. \quad (55)$$

There are two invariants, which are conserved when the electric and magnetic field components are transformed from the one inertial frame of reference to another:

$$\mathbf{E'H'} = \mathbf{EH} \quad (56)$$

and

$$E'^2 - H'^2 = E^2 - H^2. \quad (57)$$

The transformations (55) conserve the first invariant (56). The second invariant (57) is conserved when

$$\frac{1}{c} - \frac{c}{S^2} \to 0. \quad (58)$$

We can introduce a dimensionless parameter $\delta$ such as

$$\frac{1}{c} - \frac{c}{S^2} = \frac{\delta}{c}. \quad (59)$$

The values of parameter $\delta$ should lie in the semi-open interval [0, 1): when $\delta = 0$, then $S = c$, and when $\delta \to 1$, then $S \to \infty$. The ratio $S/c$ is related to the parameter $\delta$ by

$$\frac{S}{c} = \frac{1}{\sqrt{1-\delta}}, \quad (60)$$

e. g. $S = 2 \cdot c$ when $\delta = 0.75$. The question of the value of parameter $\delta$ is opened and is strongly correlated with precision of the measurements of the electromagnetic processes.

## V. EXPERIMENTAL TESTS OF THE SPECIAL THEORY OF SUPER RELATIVITY

To verify the proposed special theory of super relativity, a very precise measurement of speed of light emitted by the body moving with a constant velocity in the inertial frame of reference is needed.

The scheme of the experiment we propose is based on the measurement of the speed of light in the only one direction: from a transmitter to a receiver. The goal of this experiment is to compare the speed of light emitted by the transmitter moving toward receiver with the speed of light emitted by the transmitter moving off receiver.

In Section III.B we considered the case when the light propagates in empty space in forward and backward directions in the inertial system $K'$ moving with velocity $v = 1.6 \cdot 10^4$ m/s with respect to another inertial system $K$. There we obtained the formulas (27) and (28) for the values of the speed of light emitted the opposite directions and the ratio (29) of these values.

The velocity of $1.6 \cdot 10^4$ m/s could be obtained in space near the Earth in the relative movement of two artificial earth satellites each of those moves with an orbital velocity $v = 8000$ m/s at the altitude of $4 \cdot 10^5$ m above the ground. In this case we can emit a laser ray of light from the satellite-transceiver to the satellite-receiver for a distance of $4.0333 \cdot 10^6$ m when we assume the minimal distance from the laser ray trajectory to the ground to be $10^5$ m.

According to (27) and (60), when the transmitter moves toward the receiver (see Fig. 2 (a)), the laser ray of light emitted from the transmitter to the receiver passes this distance for a time interval either 13.4529 or 13.4536 ms, if we assume the value of parameter $\delta$ equal to either 1 or 0, respectively.

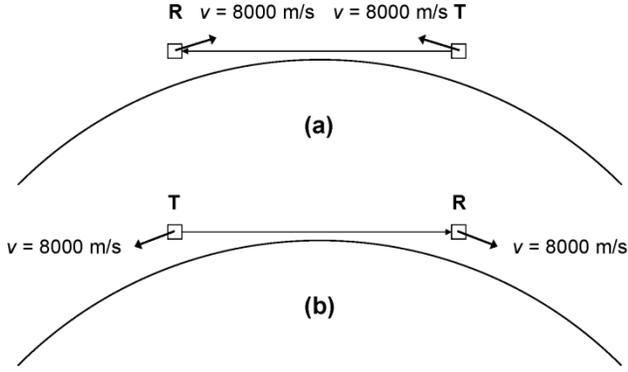

Fig. 2. Emission of the laser rays of light from the earth satellite-transmitter (T) to the earth satellite-receiver (R): (a) the transmitter moves toward receiver; (b) the transmitter moves off receiver. Both transmitter and receiver move around the Earth at the altitude $4\cdot10^5$ m with the velocities 8000 m/s. One quarter of the Earth circumference is shown.

According to (28) and (60), when the transmitter moves off receiver (see Fig. 2 (b)), the laser ray of light emitted from the transmitter to the receiver passes this distance for a time interval either 13.4544 or 13.4536 ms, if we assume the value of parameter $\delta$ equal to either 1 or 0, respectively.

Thus, according to the proposed special theory of super relativity, the difference in those two time intervals is around $1.5 \cdot 10^{-6}$ s when $S \to \infty$, and $1.1 \cdot 10^{-6}$ s when $S = 2 \cdot c$. To measure this time difference we have to know the positions of both satellites with accuracy better than 450 m and synchronize the clocks of both satellites with a precision better than $10^{-7}$ s. Despite the synchronization of the clocks could be done at the ground before launching the satellites, to register so small time intervals and to measure the positions of the satellite from the ground with so high precision is much challenged task.

Why do we need to measure the speed of light in the only direction of propagation of the light from transmitter to receiver and should not use any reflectors and multi-passes of the light between transmitter and receiver?

In any experiments of A. A. Michelson and E. W. Morley type [4]–[6] the measured time is the mean time of passing the light from the transmitter to receiver ($t_f$) and back ($t_b$):

$$\langle t \rangle = (t_f + t_b)/2. \quad (61)$$

To calculate the time intervals $t_f$, we use equations (11). The time intervals $t_b$ can be expressed by

$$t_{xb} = \gamma\left(t' - x'\frac{v}{S^2}\right), \quad t_{yb} = t', \quad t_{zb} = t', \quad (62)$$

thus,

$$\langle t_x \rangle = \gamma t', \quad \langle t_y \rangle = t', \quad \langle t_z \rangle = t'. \quad (63)$$

If in equations (63) we assume the velocity of super light $S$ equal to the velocity of propagation of the electromagnetic interaction in empty space $c$, and the velocity of the moving frame of reference $v$ equal to the orbital velocity of the Earth, we obtain

$$\langle t_x \rangle / \langle t_y \rangle = \langle t_x \rangle / \langle t_z \rangle \approx 1.000000005, \quad (64)$$

that is rather difficult to measure.

## VI. SUPERLUMINAL NEUTRINOS AND CONSTRAINTS FROM THE MUON MASS MEASUREMENTS

Six years after this study was completed the OPERA collaboration reported an anomaly [3] in the ratio

$$r = (S - c)/c \quad (65)$$

of the neutrino velocity $S$ to the speed of light in free space $c$ of

$$r = (2.48 \pm 0.28(stat.) \pm 0.30(sys.)) \times 10^{-5}. \quad (66)$$

Consider the decay of the muon at rest into an electron, a muon neutrino and an electron antineutrino. Let electron in the decay has the energy of

$$E_e = m_\mu/2 \quad (67)$$

expressed in *natural units*, where the speed of light $c = 1$. Here we use the known value [7] of the muon mass

$$m_\mu = 105.6583668 \, MeV. \quad (68)$$

The electron momentum can be obtained from the conventional expression:

$$p_e = (E_e^2 - m_e^2)^{\frac{1}{2}}, \quad (69)$$

where $m_e$ is the electron mass. The value of the electron velocity $\beta_e$ we calculate from the equation as follows:

$$\beta_e = p_e/E_e = 0.999953219. \quad (70)$$

Assuming the neutrino speed measured by the OPERA collaboration [3] to be the superluminal velocity $S$, we can represent it using Eq. (65) and equality of $c = 1$ in the form:

$$S = 1 + r, \quad (71)$$

where $r$ is determined by (66). Then we can rewrite the expression for the electron energy in the modified relativistic kinematics described in Section III in the form as follows:

$$E_e^S = m_e(1+r)^2/(1 - \beta_e^2/(1+r)^2)^{\frac{1}{2}}. \quad (72)$$

Here $\beta_e$ is the measured electron velocity (70).

Taking into account that the electron energy is a half of the muon energy at rest, we can obtain the muon mass in the modified relativistic kinematics:

$$m_\mu^S = 2E_e^S/(1+r)^2 = 85.4177895 \, MeV, \quad (73)$$

that is 19.16% less than the known value of the muon mass (68) precisely measured in the hyperfine Zeeman transitions in the ground state of muonium in a strong magnetic field at LAMPF [8], [9]. Using *SI units* gives the same result. This completely excludes introducing the superluminal velocity $S$ into special theory of relativity.

## VII. Conclusions

The special theory of super relativity is based on gentle modifications of A. Einstein's special relativity. Replacing in Lorentz transformations the speed of light by superluminal velocity changes the relativistic kinematics and breaks the invariance of two Maxwell's equations in the inertial frames. In addition, even a small distinction of superluminal velocity from the speed of light, as measured by the OPERA collaboration, leads in the modified kinematics to low value of the muon mass reconstructed from the muon decay products. This value contradicts to the muon mass measurements performed in the hyperfine Zeeman transitions in the ground state of muonium in a strong magnetic field at LAMPF. That strongly excludes introducing the superluminal velocity into special theory of relativity.

## VIII. Acknowledgment

The author thanks Prof. L. I. Sarycheva of SINP MSU (Moscow, Russia) for prolific discussions.